\documentclass[12pt]{article}
\usepackage{graphicx}
\usepackage{latexsym}
\usepackage{amscd}
\usepackage[utf8]{inputenc}
\usepackage[english,russian]{babel}
\usepackage{amsmath}
\usepackage{amssymb}
\usepackage{indentfirst}
\usepackage[small,centerlast]{caption2}
\usepackage{longtable}
\usepackage[dvipsnames]{color}
\usepackage{cite}

\def\red#1{\textcolor{red}{#1}}

\let\vec=\mathbf

\makeatletter
\renewcommand{\section}{\@startsection {section}{1}{\z@}%
	{-3.5ex \@plus -1ex \@minus -.2ex}%
	{2.3ex \@plus.2ex}%
	{\normalfont\Large}}
\renewcommand{\subsection}{\@startsection{subsection}{2}{\z@}%
	{-3.25ex\@plus -1ex \@minus -.2ex}%
	{1.5ex \@plus .2ex}%
	{\normalfont\large\itshape}}
\renewcommand{\subsubsection}{\@startsection{subsubsection}{3}{1em}%
	{-3.25ex\@plus -1ex \@minus -.2ex}%
	{-1.5em \@plus .2em}%
	{\normalfont\normalsize\bfseries}}

\makeatother

\begin{document}
\centerline{\textbf{\Large ДИНАМИКА ПЛАЗМЫ}}

\noindent\textit{УДК  533.9}

\begin{center}
	
	{\Large \textbf{DIFFUSION IN PLASMA: THE HALL EFFECT, COMPOSITIONAL WAVES, 
			AND CHEMICAL SPOTS\\}}
	
	\medskip
	
	{\large \textbf{V.~Urpin}}
	
	\medskip
	
	\textit{A.F.Ioffe Institute of Physics and Technology, 
		St. Petersburg, Russia} \\
		\textit{INAF, Osservatorio Astrofisico di Catania, Catania, Italy}\\[10pt]
\end{center}

\renewcommand{\abstractname}{}
\begin{abstract}
	\noindent\small We consider diffusion caused by a combined influence 
	of the electric 
	current and Hall effect, and argue that such diffusion can form 
	inhomogeneities of a chemical composition in plasma. The considered 
	mechanism can be responsible for a formation of element spots in 
	laboratory and astrophysical plasmas. This current-driven diffusion 
	can be accompanied by propagation of a particular type of waves in which 
	the impurity number density oscillates alone. These compositional
	waves exist if the magnetic pressure in plasma is much greater 
	than the gas pressure.
\end{abstract}

\section{INTRODUCTION}

Often laboratory and astrophysical plasmas are multicomponent, and
diffusion plays an important role in many phenomena in such plasmas. 
For instance, diffusion can be responsible for the formation of chemical 
inhomogeneities which influence emission, heat transport, conductivity, 
etc (see, e.\:g., [1--3]). In thermonuclear fusion 
experiments, the source of trace elements is usually the chamber walls, 
and diffusion determines the penetration depth of these elements and 
their distribution in plasma (see, e.\:g., [4--6]).
Even a small admixture of heavy ions increases drastically radiative 
losses of plasma and changes its thermal properties. In astrophysical 
conditions, diffusion leads to the formation of element spots detected 
on the surface of many stars (see, e.\:g., [7--9]).
Usually, diffusion in astrophysical bodies is influenced by a number 
of factors such as gravity, radiative force, magnetic field, 
temperature gradient, etc. (see, e.\:g., [10]). Under such 
conditions, diffusion processes may exhibit some rather unexpected 
properties that still have not been studied in laboratories. 

Diffusion in plasma can differ qualitatively from that in neutral 
gases because of the presence of electrons and electric currents. 
A mean motion of electrons caused by electric currents provides an 
additional internal force that results in diffusion of trace elements 
(see, e.\:g., [11]). One more important contribution of electrons 
in diffusion is relevant to the Hall effect. The magnetic field can 
magnetize the charged particles that leads to the anisotropic transport. 
In the case of electron transport, such anisotropy is characterized 
by the Hall parameter, $x_e = \omega_{Be} \tau_e$, where $\omega_{Be} = 
e B/m_e c$ is the gyrofrequency of electrons and $\tau_e $ is their 
relaxation time, $B$ is the magnetic field. In a hydrogen plasma, 
$\tau_e = 3 \sqrt{m_e} (k_b T)^{3/2}/4 \sqrt{2 \pi} e^4 n_e \Lambda $ 
(see, e.\:g., [10]) where $n_e$ and $T$ are the number density 
of electrons and their temperature, respectively, $\Lambda$ is the 
Coulomb logarithm. At $x_e \geq 1$, the rates of diffusion along 
and across the magnetic field become different and, in general, 
diffusion can lead to the inhomogeneous distribution of elements.

In this paper, we consider one more diffusion process that can 
lead to formation of chemical inhomogeneities in plasma. This 
process is caused by 
the combined influence of the electric currents 
and the Hall effect. Using a simple model, we show that the 
interaction of the electric current with trace elements leads to 
their diffusion in the direction perpendicular to both the 
electric current and the magnetic field. This type of diffusion can 
alter the distribution of chemical elements in plasma and 
contribute to the formation of chemical spots even if the magnetic 
field is relatively weak and does not magnetize electrons ($x_e 
\ll 1$). We also argue that the current-driven diffusion in 
combination with the Hall effect can be the reason of the 
particular type of modes in which the number density of a trace 
element oscillates alone.      

\section{BASIC EQUATIONS AND DIFFUSION COEFFICIENTS}

Consider plasma with the magnetic field parallel to the axis 
$z$, $\vec{B} = B \vec{e}_{z}$, where $(s, \varphi, z)$  are 
cylindrical coordinates and $(\vec{e}_s, \vec{e}_{\varphi}, 
\vec{e}_{z})$ \red{are} the corresponding unit vectors, respectively. 
We assume that plasma is cylindrical and the magnetic field 
depends on the cylindrical radius alone, $B=B(s)$. Then, the 
electric current is given by 
\begin{equation}
j_{\varphi} = - (c/4 \pi) \partial B/\partial s.  
\end{equation}
We suppose that $j_{\varphi} \rightarrow 0$ at large $s$ and,
hence, $B \rightarrow B_0=\text{const}$ at $s \rightarrow \infty$.
Note that the dependence $B(s)$ can not be an arbitrary function 
of $s$ because, generally, the cylindrical magnetic configurations 
are unstable if $B(s)$ increases with $s$ or decreases sufficiently 
slowly (see, e.\:g., [12--14]). In astrophysical 
bodies, the magnetic field usually has a more complex topology 
than our simple configuration. However, this model describes 
correctly the main qualitative features of current-driven diffusion. 
In some cases, this model can even mimic the magnetic field 
in certain regions of a star. For example, the field near the 
magnetic pole has a topology very close to our model (1) (see, 
e.\:g., [15]).    

We assume that plasma consists of electrons $e$, protons $p$, 
and a small admixture of heavy ions $i$. The number density 
of species $i$ is small and it does not influence dynamics 
of plasma. Therefore, this species can be treated as trace 
particles that interact only with a background hydrogen 
plasma. The partial momentum equations in fully ionized multicomponent 
plasma has been considered by a number of authors (see, e.\:g.,
[16, 17]). These equations can be obtained by 
multiplying the Boltzmann kinetic equation for each species by 
its velocity and integrating over it. The momentum equation for 
particles $\alpha$ ($\alpha = e, p, i$) reads
\begin{eqnarray}
m_{\alpha} n_{\alpha} \left[
\dot{{\bf V}}_{\alpha}
+ ({\bf V}_{\alpha} \cdot \nabla) {\bf V}_{\alpha} \right]
= - \nabla p_{\alpha} 
+ n_{\alpha} {\bf F_{\alpha}} +
\nonumber \\
e Z_{\alpha} n_{\alpha} \left({\bf E} + \frac{{\bf V}_{\alpha}}{c}
\times {\bf B} \right) + {\bf R}_{\alpha},
\end{eqnarray}  
the dot denotes the partial time derivative. 
Here, $m_{\alpha}$ and $Z_{\alpha}$ are the mass and the charge number
of particles $\alpha$, $n_{\alpha}$ and $p_{\alpha}$ are their 
number density and pressure, respectively, ${\bf V}_{\alpha}$ is 
the mean velocity, ${\bf F}_{\alpha}$ is an external force acting 
on the particles $\alpha$; ${\bf E}$ and ${\bf B}$ are the 
electric and magnetic fields, respectively; ${\bf R}_{\alpha}$ 
is the internal friction force caused by the collisions of particles 
$\alpha$ with other sorts of particles. Since ${\bf R}_{\alpha}$ 
is the internal force, the sum of ${\bf R}_{\alpha}$ over $\alpha$ 
is zero in accordance with the Newton's third law. Usually, the force
${\bf F}_{\alpha}$ is the sum of the gravitational and radiation
force. Below we will neglect it.

If there are no mean hydrodynamic velocity and only diffusive 
velocities of trace elements are non-vanishing, the partial 
momentum equation for particles $\alpha$ reads
\begin{equation}
- \nabla p_{\alpha} + Z_{\alpha} e n_{\alpha} \left( {\bf E} + 
\frac{{\bf V}_{\alpha}}{c} 
\times {\bf B} \right) + {\bf R}_{\alpha} 
= 0. 
\end{equation}
In Eq. (3) for the trace particles $i$, we can represent the 
friction forces ${\bf R}_i$ as ${\bf R}_i = {\bf R}_{ie} + 
{\bf R}_{ip}$, where the force $\vec{R}_{ie}$ is caused by 
scattering of the ions $i$ on the electrons and $\vec{R}_{ip}$ 
by scattering on the protons. 

If $n_i$ is small compared to the number density of protons 
$n_p$, $\vec{R}_{ie}$ is given approximately by      
\begin{equation}
\vec{R}_{ie} = - \frac{Z_i^2 n_i}{n_p} \vec{R}_{e}
\end{equation}
where $\vec{R}_{e}$ is the force acting on the electron
gas (see, e.\:g., [18]). Since $n_i \ll n_p $, 
$\vec{R}_{e}$ is determined mainly by scattering of electrons 
on protons but scattering on ions $i$ gives a small 
contribution. Therefore, we can use for $\vec{R}_{e}$ the 
expression for one component hydrogen plasma calculated by 
Braginskii [16]. In our model of a cylindrical plasma 
configuration, this expression reads
\begin{equation}
\vec{R}_{e} = - \alpha_{\perp} \vec{u} + \alpha_{\wedge} \vec{b}
\times \vec{u} - \beta^{uT}_{\perp} \nabla T - \beta^{uT}_{\wedge}
\vec{b} \times \nabla T,
\end{equation}   
where $\vec{u} = - \vec{j}/en$ is the difference between the mean 
velocities of electrons and protons; $\alpha_{\perp}$, 
$\alpha_{\wedge}$, $\beta^{uT}_{\perp}$, and $\beta^{uT}_{\wedge}$ 
are the coefficients calculated 
in [16]; $\vec{b} = 
\vec{B}/B$. The first two terms on the r.h.s. of Eq.(5) describe the
standard friction force caused by a relative motion of the electron 
and proton gases. The last two terms on the r.h.s. of Eq.(5)
represent the so-called thermoforce caused by a temperature 
gradient. This part of $\vec{R}_{e}$ is responsible for 
thermodiffusion. For the sake of simplicity, we consider plasma
with a uniform temperature, $\nabla T =0$.


Taking into account that $\vec{u} = u \vec{e}_{\varphi}$ in our model 
and using coefficients $\alpha_{\perp}$ and $\alpha_{\wedge}$   
calculated 
in [16], we obtain the followng expressions for 
the cylindrical components of ${\bf R}_{ie}$
\begin{equation}
R_{ie \varphi} = Z_i^2 n_i \left( \frac{m_e}{\tau_e} \delta_1 u
\right), \;\,
R_{ie s} = Z_i^2 n_i \left( \frac{m_e}{\tau_e} \delta_4 u
\right), 
\end{equation}
where
\begin{eqnarray}
\delta_1 \! = \! 1 \!-\! \delta_3^{-1} (1.84 + 6.42 x^2), 
\delta_4 \! = \! \delta_3^{-1} x (0.78 + 1.7 x^2),  \nonumber \\
\delta_3 = x^4 + 14.79 x^2 + 3.77, \;\;\ x = \omega_{Be} \tau_e.
\end{eqnarray}

The force $\vec{R}_{ip}$ consists of two parts as well, $\vec{R}^{'}_{ip}$
and $\vec{R}^{''}_{ip}$, which are proportional to the relative 
velocity of ions $i$ and protons and to the temperature gradient,
respectively. The thermoforce is vanishing in our model. The 
friction force $\vec{R}^{'}_{ip}$ can be easily calculated in the most 
interesting case when the mass of a species $i$, $m_i$, is greater 
than the proton mass, $m_p$. In this case, $\vec{R}^{'}_{ip}$ is 
proportional to the relative velocity of heavy ions and the background 
plasma, $(\vec{V}_p - \vec{V}_i)$. Taking into account that the mean 
velocity of the background plasma in our simplified model is assumed 
to be zero, the friction force can be represented as (see, e.\:g., 
[1, 17]) 
\begin{equation}
\vec{R}^{'}_{ip} = \frac{0.42 m_i n_i Z^2_i}{\tau_i} (-\vec{V}_i),
\end{equation} 
where $\tau_i = 3\sqrt{m_i} (k_B T)^{3/2} / 4 \sqrt{2 \pi} e^4 n_p 
\Lambda$ and $\tau_{i}/ Z^2_i$ is the characteristic timescale of 
ion-proton scattering; we assume that Coulomb logarithms are the 
same for all types of scattering. 
Since the number density of trace particles is small,
we can suppose in calculations $n_p \approx n_e = n$. 



The momentun equation for the species $i$ (see Eq.(3)) contains 
cylindrical components of the electric field, $E_s$ and $E_{\varphi}$.
These components can be determined from the momentum equations (3)
for electrons and protons
\begin{eqnarray}
- \nabla (n_e k_B T) - e n_e \left( \vec{E} + \frac{\vec{u}}{c} \times
\vec{B} \right) + \vec{R}_e = 0, \\
- \nabla (n_p k_B T) + e n_p \vec{E} - \vec{R}_e + \vec{F}_p = 0.  
\end{eqnarray}

Taking into account the condition of hydrostatic equilibrium
and quasi-neutrality ($n_e \approx n_p$), we obtain the
following expressions for the radial and azimuthal electric fields
\begin{equation}
E_s = - \frac{uB}{2c} - \frac{1}{e} \left( \frac{m_e u}{\tau_e} \delta_4
\right) , \;\;
E_{\varphi} = - \frac{1}{e} \left( \frac{m_e u}{\tau_e} \delta_1 \right) .
\end{equation}
Substituting Eqs. (6), (8), and (11) into Eq. (3) for the trace particles $i$, 
we arrive to the expression for a diffusion velocity ${\bf V}_i$, 
\begin{equation}
\vec{V}_i = V_{is} \vec{e}_s + V_{i \varphi} \vec{e}_{\varphi}, 
\;\;\;\;\; V_{is} = V_{n_i} + V_B,
\end{equation} 
where
\begin{equation}
V_{n_i} = - D \frac{d \ln n_i}{d s}, \quad 
V_B =D_B \frac{d \ln B}{d s}, \quad
V_{i \varphi} = D_{B \varphi} \frac{d B}{d s}; 
\end{equation}
$V_{n_i}$ is the velocities of ordinary diffusion and $V_B$ 
and $V_{i \varphi}$ are the radial and azimuthal diffusion velocities 
caused by the electric current. The corresponding diffusion 
coefficients are
\begin{eqnarray}
D = \frac{2.4 c_i^2 \tau_i}{Z_i^2 (1 + q^2)} , \;\;\; c_i^2 = \frac{k_B T}{m_i},
\;\;\; q = \frac{2.4 e B \tau_i}{Z_i m_i c},  \nonumber \\
D_B \!=\! \! \frac{2.4 c B \sqrt{m_e/m_i}}{4 \pi e n (1 \! +  \! q^2)} \! \!  
\left[  \left( 1 \!-\! \frac{1}{Z_i} \right) 
(\delta_4 + q \delta_1 ) \!-\! \frac{x}{2 Z_i} \! \right], \nonumber \\
D_{B \varphi} \!=\! \! \frac{2.4 c \sqrt{m_e/m_i}}{4 \pi e n 
(1 \! +  \! q^2)} \! \!  
\! \left[  \left( 1 \!-\! \frac{1}{Z_i} \right) 
(\delta_1 - q \delta_4 ) \!+\! \frac{q x}{2 Z_i} \! \right].
\end{eqnarray} 
Eqs.~(12)--(14) describe the drift of ions $i$ under a combined 
influence of $\nabla n_i$ and $\vec{j}$. 

If magnetic field is weak and $x \ll 1$, 
Eq. (14) yields
\begin{eqnarray}
D \approx \frac{2.4 c_i^2 \tau_i}{Z_i^2}, \;\;\; 
D_B \approx \frac{2.4 c_A^2 \tau_i}{Z_i A_i} (0.2 Z_i - 0.7), \nonumber \\
D_{B \varphi} \approx 1.2 \sqrt{\frac{m_e}{m_i}} 
\frac{c (Z_i - 1)}{4 \pi en Z_i}. 
\end{eqnarray}
where $c_A^2 = B^2 / (4 \pi n m_p)$ [18]. 
The coefficient $D$ is always positive but two other diffusion 
coefficients can be positive or negative depending on the 
parameters of plasma.  

In the opposite case of a very strong magnetic field, $q \gg 1$,
Eq. (14) yields
\begin{equation}
D \approx \frac{2.4 c_i^2 \tau_i}{Z_i^2 q^2}, \;\;\; 
D_B \approx - \frac{1.2 c_A^2 \tau_i}{Z_i A_i q^2} , \;\;\;
D_{B \varphi} \approx \frac{c}{8 \pi en}. 
\end{equation}
In our model, diffusion in the radial direction is strongly
suppressed because all sorts of particles are magnetized.  
For instance, the coefficient $D$ (Eq. (16)) and the corresponding 
diffusion velocity $V_{ni}$ which characterize the standard
diffusion in the $s$-direction are $\approx q^2 \gg 1$ times 
smaller than those in the case of a weak magnetic field.
The coefficient $D_B$ is also approximately $q^2$ times 
smaller in a strong magnetic field. Note that $D_B$ reaches 
saturation and do not depend on the field strength at $q \gg 1$. 
If the electric current is fixed ($dB/ds=\text{const}$), the radial 
diffusion velocity $V_B$ caused by currents decreases $\propto 
1/B$. As far as the azimuthal diffusion is concerned, the 
coefficient $D_{B \varphi}$ does not depend on the magnetic field 
in both cases, strong and weak magnetic fields. However, 
$D_{B \varphi}$ in a strong field is greater by a factor $\sim 
\sqrt{m_i/m_e}$. 

\section{ELEMENT SPOTS CAUSED BY ELECTRIC CURRENTS}

It is generally believed that standard diffusion smoothes 
chemical inhomogeneities on a diffusion timescale $ \sim L^2/ D$ 
where $L$ is the lengthscale of a non-uniformity. This is not 
the case, however, for diffusion given by Eq.~(12). In this case, 
chemical inhomogeneities can exist during a much longer time than 
$\sim L^2/D$ because the equilibrium distribution is reached due 
to balance of two diffusion processes, standard ($\propto \nabla n_i$) 
and current-driven ($\propto dB/ds$) ones, which push ions in the 
opposite directions. As a result, $V_{is} = 0$ in the equilibrium 
state and this state can be maintained as long as the electric 
current exists. 

Note that the radial velocity is vanishing in the equilibrium state 
but the azimuthal velocity is non-zero. It turns out that impurities 
rotate around the magnetic axis even if equilibrium is reached. The 
direction of rotation depends on the sign of $dB/ds$ and is opposite 
to the electric current. Since electrons move in the same direction, 
heavy ions turn out to be carried along the flow of electrons. Different 
ions move with different velocities around the axis. If the magnetic 
field is weak ($x \ll 1$), the difference between different sorts of 
ions, $\Delta V_{i \varphi}$, is of the order of 
\begin{equation}
\Delta V_{i \varphi} \sim \frac{c}{4 \pi e n} \sqrt{\frac{m_e}{m_i}}
\frac{dB}{ds} \sim 3 \times 10^{-3} \frac{B_4}{n_{14} L_{10} A_i^{1/2}}
\;\; \frac{{\rm cm}}{{\rm s}},
\end{equation} 
where $B_4 =B /10^4$ G, $n_{14} = n/10^{14}$~cm$^{-3}$, and
$L_{10} = L /10^{10}$~cm. Since different impurities rotate around 
the magbetic axis with different velocities, periods of such 
rotation are also 
different for different ions. The difference
in periods can be estimated as  
\begin{equation}
\Delta P = \frac{2 \pi L}{\Delta V} \sim 10^6 
\frac{L_{10}^2 n_{14} A_i^{1/2}}{B_4} \;\; {\rm yrs}.
\end{equation}
If the distribution of impurities is non-axisymmetric then such
diffusion in the azimuthal direction should lead to slow 
variations in the abundance peculiarities. 
Note that in the case of a strong field ($q \gg 1$), all sorts
of trace particles rotate around the axis with the same period
that depends only on the number density and electric current.  

The condition of hydrostatic equilibrium in our model is given by
\begin{equation}
- \nabla p + \vec{j} \times \vec{B} / c = 0,
\end{equation}
where $p$ and $\rho$ are the pressure and density, respectively. 
Since the background plasma is hydrogen, $p \approx 2n k_B T$ 
where $k_B$ is the Boltzmann constant. Integrating the $s$-component 
of Eq.~(19) and taking into account that the temperature is constant
in our model, we obtain 
\begin{equation}
n = n_0 \left( 1 + \beta_0^{-1} - \beta^{-1} \right),
\end{equation}
where $\beta = 8 \pi p_0 / B^2$; $(p_0, n_0, T_0, \beta_0)$ are 
the values of $(p, n, T, \beta)$ at $s \rightarrow \infty$.

Consider the equilibrium distribution of trace elements in cylindrical 
plasma. In equilibrium, we have $V_{i s} = 0$ and Eq. (12) yields 
\begin{equation}
\frac{d \ln n_i}{d s} =
\frac{D_B}{D} \frac{d \ln B}{d s}. 
\end{equation}
The term on the r.h.s. describes the effect of electric currents 
on the distribution of trace elements. Note that this type of 
diffusion is driven by the electric current rather than an 
inhomogeneity of the magnetic field. Ocasionally, the conditions 
$d B/ ds \neq 0$ and $j \neq 0$ are equivalent in our simplified 
model. 

First we consider
the case of a weak
magnetic field with $x \ll 1$. Then, one has from Eq.~(19) 
\begin{equation}
\frac{d}{ds}(n k_B T) = - \frac{B}{8 \pi} \frac{d B}{d s}. 
\end{equation}  
Substituting Eq.~(22) into Eq. (21) and integrating, 
we obtain
\begin{equation}
\frac{n_i}{n_{i0}} = \left( \frac{n}{n_0} \right)^{\mu},
\end{equation}
where
\begin{equation}
\mu =  - 2 Z_i (0.2 Z_i - 0.7)
\end{equation}
and $n_{i0}$ is the value of $n_i$ at $s \rightarrow \infty$. 
Denoting the local abundance of the element $i$ as $\gamma_i = 
n_i/n$ and taking into account Eq.~(19), we have 
\begin{equation}
\frac{\gamma_i}{\gamma_{i0}} =
\left( \frac{n}{n_0} \right)^{\mu-1} = 
\left( 1 + \frac{1}{\beta_0} - \frac{1}{\beta} \right)^{\mu -1},
\end{equation}
where $\gamma_{i0} = n_{i0}/n_0$. 
Local abundances turn out to be flexible to the field strength 
and, particularly, this concerns the ions with large charge
numbers. If other mechanisms of diffusion are negligible and 
the distribution of elements is basically current-driven, then
the exponent $(\mu -1)$ can reach large negative values for 
elements with large $Z_i$ and, hence, produce strong abundance 
anormalies. For instance, $(\mu -1)$ is equal 1.16, $-0.52$, and 
$-2.04$ for $Z_i=2$, 3, and 4, respectively. Note that $(\mu -1)$ 
changes its sign as $Z_i$ increases: $(\mu -1) >0$ if $Z_i = 2$ 
but $(\mu -1)<0$ for $Z_i \geq 3$. Therefore, elements with $Z_i 
\geq 3$ are in deficit ($\gamma_i < \gamma_{i0}$) in the region 
with a weak magnetic field ($B < B_0$) but, on the contrary, 
these elements should be overabundant in the region where 
the magnetic field is stronger than $B_0$. 

The distribution of the impurities can be substantially different
if the magnetic field is strong and $q \gg 1$. Using the same 
procedure as in the case of a weak field, we obtain
\begin{equation}
\frac{\gamma_i}{\gamma_{i0}} =
\left( \frac{n}{n_0} \right)^{Z_i-1} = 
\left( 1 + \frac{1}{\beta_0} - \frac{1}{\beta} \right)^{Z_i -1}.
\end{equation}
Therefore, all trace elements with $Z_i > 1$ are overabundant
in the regions with the magnetic field weaker than $B_0$. On
the contrary, these elements are underabundant in the regions
with a stronger nagnetic field.

Note that calculating $\vec{E}$ from Eqs.~(9) and (10), we 
neglect the electric field generated by 
redistribution of 
heavy ions because the number density of such ions is small.
This electric field will decrease 
formation of spots and 
can produce departures from the simple picture outlined in 
this section. However, these departures are basically small
since $n_i \ll n$, and they begin to play an important role
only if the electric field generated by
the redistribution of 
impurities in the spot is comparable to $\vec{E}$. Using 
Eqs.~(9) and (10), one can 
estimate that the influences 
of these electric fields becomes comparable if $Z_i n_i \sim 
n$ in the spot. This equation determines the impurity number 
density above which our consideration is unjustified.     
 
\section{COMPOSITIONAL WAVES}

In our simplified model of plasma cylinder with the velocity
given by Eq. (12), the continuity equation for trace ions $i$ 
reads
\begin{equation}
\frac{\partial n_i}{\partial t} + \frac{1}{s} \frac{\partial}{\partial s}
\left( s n_i V_{is} \right) + \frac{1}{s} \frac{\partial}{\partial
\varphi} (n_i V_{i \varphi} )  = 0.
\end{equation}
Consider the behavior of small disturbances of the number 
density of trace ions by making use of a linear 
analysis of Eq.~(27).  In the basic (unperturbed) state, 
plasma is assumed to be in a diffusive equilibrium and, hence, 
the unperturbed impurity number density satisfies Eq.~(21).
Since the number density of impurity $i$ is small, its 
influence on parameters 
in the basic state is negligible. 
For the sake of simplicity, we consider disturbances that 
do not depend on $z$. Denoting disturbances of the impurity 
number density by $\delta n_i$ and linearizing Eq. (26), we 
obtain the equation governing the evolution of such small 
disturbances,   
\begin{eqnarray}
\frac{\partial \delta n_i}{\partial t} - \frac{1}{s} 
\frac{\partial}{\partial s}
\left( s D \frac{\partial \delta n_i}{\partial s} - s \delta n_i
\frac{D_B}{B} \frac{dB}{ds} \right) +
\nonumber \\
\frac{1}{s} \frac{\partial}{\partial \varphi}
\left( \delta n_i D_{B \varphi} \frac{d B}{d s} \right)= 0.
\end{eqnarray}
For the purpose of illustration, we consider only 
disturbances with the wavelengths shorter than the lengthscale
of unperturbed quantities. In this case, we can use the so 
called local approximation for a consideration of linear
waves and assume that small disturbances
are $\propto \exp(-i k s - i M \varphi)$ where $k$ is the 
wavevector ($ks \gg 1$) and $M$ is the azimuthal wavenumber. 
Since the basic state does not depend on time, $\delta
n_i$ can be represented as $\delta n_i \propto e^{i\omega t - 
i k s - iM \varphi}$ where $\omega$ should be calculated 
from the dispersion equation. Substituting $\delta n_i$ in 
such form into Eq.~(28), we obtain the following dispersion 
equation 
\begin{eqnarray}
i \omega = - \omega_{R} + i \omega_{I}, \;\;\; \omega_{R} =
D k^2, \;\;\; \omega_{I} = \omega_S + \omega_{\varphi}, \nonumber \\
\omega_{S} = k D_B \frac{d \ln B}{ds}, \;\; 
\omega_{\varphi}= \frac{M}{s} BD_{B \varphi} \frac{d \ln B}{ds}.
\end{eqnarray}
This dispersion equation describes spiral waves in which 
only the number density of impurities oscillates and, therefore,
such waves can be called ``compositional''. The quantity 
$\omega_{R}$ characterizes decay of these waves with the 
characteristic timescale $\sim (D k^2)^{-1}$ typical for a
standard diffusion. The frequency $\omega_{I}$ describes
oscillations of impurities caused by the combined action of
electric current and the Hall effect. Note that $\omega_{I}$
can be of any sign but $\omega_{R}$ is always positive. 
The frequency $\omega_S$ characterizes oscillations in the
radial direction and $\omega_{\varphi}$ is in the azimuthal 
direction.

The compositional waves are aperiodic if $\omega_{R} >
|\omega_{I}|$ and oscillatory if $|\omega_{I}| > \omega_{R}$.
We consider the compositional waves in particular cases of
weak ($x \ll 1$) and strong ($q \gg 1$) magnetic fields.  

{\bf Weak magnetic field ($x \ll 1$).}
If $ks \gg M$ (radial waves), the condition $|\omega_{I}| > 
\omega_{R}$ in a weak field is equivalent to  
\begin{equation}
c_A^2/c_s^2 > Z_i^{-1} |0.21 Z_i - 0.71|^{-1} kL,
\end{equation}
where $L = | d \ln B/ ds|^{-1}$ and $c_s$ is the sound speed, 
$c_s^2 = k_B T/m_p$. In the opposite case $M \gg ks$ (azimuthal 
waves), the compositional waves are oscillatory if
\begin{equation}
c_A^2 / c_s^2 \gg  x (ks/M) (k L).
\end{equation}
Both conditions (30) and (31) require 
very strong
magnetic field 
so the magnetic pressure is substantially greater 
than the gas pressure. The frequency of compositional 
waves is higher in the region where the magnetic field 
has a stronger gradient or, in other words, where the 
density of electric currents is greater. Note that different 
impurities oscillate with different frequences. 

Consider first the radial waves with $M=0$. Substituting $M=0$ 
into Eq.~(29), we obtain the dispersion equation for such
waves in the form 
\begin{equation}
i \omega = - \omega_{R}  +  i \omega_{B}, \quad \omega_{R} =
D k^2, \quad \omega_{B} =  k D_B \frac{d \ln B}{ds}.
\end{equation}
This dispersion equation describes waves in which 
only the number density of trace particles oscillates and 
oscillations of $n_i$ occur only in the radial direction. 
The order of magnitude estimate of $\omega_{S}$ yields 
\begin{equation}
\omega_{I} \sim k c_A \frac{1}{Z_i A_i} \frac{c_A}{c_i}\frac{l_i}{L},
\end{equation}
where $l_i = c_i \tau_i$ is the mean free-path of ions $i$.
Note that different impurities oscillate with different 
frequences. Therefore, if there are several sorts of trace 
ions in plasma, the chemical structure should exhibit 
variations of local abundances under the influence of 
compositional waves.

The dispersion equation for non-axisymmetric waves with 
$M \gg ks$ reads in a weak field
\begin{equation}
i \omega = - \omega_{R} + i \omega_{B \varphi}, \quad 
\omega_{B \varphi} =\frac{M}{s} B D_{B \varphi} \frac{d \ln B}{ ds}.
\end{equation}
In non-axisymmetric waves, trace ions rotate around the 
cylindrical axis with the frequency $\omega_{\varphi}$ and 
decay slowly on the diffusion timescale $\sim \omega_R^{-1}$. 
The frequency of such waves is typically higher than that of 
the radial  waves. One can estimate the ratio of these 
frequencies as
\begin{equation}
\frac{\omega_{\varphi}}{\omega_S} \sim \frac{B D_{B \varphi}}{D_B}
\sim \frac{1}{A_i x}\frac{M}{ks}.
\end{equation}
Since these estimates are justified only in the case of 
a weak magnetic field ($x \ll 1$), the period of 
non-axisymmetric waves is shorter for waves with $M > 
A_i x (ks)$. The ratio of diffusion timescale and period 
of non-axisymmetric waves is 
\begin{equation}
\frac{\omega_{\varphi}}{ \omega_R} \sim \frac{1}{ x} 
\frac{c_A^2} {c_s^2} \frac{Z_i}{A_i} \frac{1}{kL}
\end{equation} 
and can be large. Therefore, azimuthal waves can be oscillatory 
as well.

{\bf Strong magnetic field ($q \gg 1$)}
In a strong magnetic field, the order of magnitude estimates
of the characteristic frequencies are   
\begin{equation}
\omega_S \approx \frac{k}{2 q} \frac{j_{\varphi}}{e n},
\;\;\;\;  \omega_{\varphi} \approx - \frac{M}{2s} 
\frac{j_{\varphi}}{en}.
\end{equation}
Like the case of a weak field, the frequency of compositional 
waves is higher in the region where the density of the electric 
currents is greater. Oscillations of different trace ions 
occur with different frequencies in radial waves but azimuthal
oscillations have the same frequency for different impurities. 
The frequency of azimuthal waves is higher 
than that of radial waves if
\begin{equation}
M \gg \frac{ks}{q}.
\end{equation}
If the magnetic field is 
\red{so} strong that $q \gg 1$ than 
the azimuthal waves oscillate with a higher frequency 
than the radial ones even for not very large $M$. The 
condition that radial waves exists in a strong magnetic 
field, $|\omega_S| \gg \omega_R$, is given by  
\begin{equation}
\frac{c_A^2}{c_s^2} > \frac{2}{Z_i} kL.
\end{equation}
Similar to the case of a weak magnetic field, compositional 
waves occur in plasma only if the magnetic pressure is
greater than the gas 
pressure. The analogous condition for 
azimuthal waves, $\omega_{\varphi} \gg \omega_R$, reads  
\begin{equation}
\frac{c_A^2}{c_s^2} > \frac{1}{q Z_i}\frac{ks}{M} kL.
\end{equation}
Note that this condition can be satisfied even if the
magnetic pressure is smaller than the gas one but $q$
and $M$ are large. 

\section{CONCLUSIONS}

We have considered diffusion of heavy ions under the influence 
of electric currents. Generally, the diffusion velocity in this 
case can be comparable to or even greater than that caused 
by other diffusion mechanisms. The current-driven diffusion 
can form chemical inhomogeneities even if the magnetic field 
is relatively weak whereas other diffusion mechanisms require 
a substantially stronger magnetic field.  

The current-driven diffusion is relevant to the Hall effect and,
therefore, it leads to a drift of ions in the direction perpendicular 
to both the magnetic field and the electric current. As a result, 
distribution of chemical elements in plasma depends essentially on 
the geometry of the magnetic fields and the electric current. Chemical 
inhomogeneities can manifest themselves, for example, by emission 
in spectral lines and a non-uniform plasma temperature. Usually, 
diffusion processes play an important role in plasma if hydrodynamic
motions are very slow. In some cases, however, chemical spots can
be formed even in flows with a relatively large velocity but with
some particular topology (for example, a rotating flow). This can 
occur usually in laminar flows. Unfortunately, such flows often are 
unstable in magnetized plasma. This is particularly concerned 
to the flows with a large Hall parameter since hydrodynamic motions in
such plasma typically are unstable even in the presence of a weak 
shear (see, e.\:g., [19--21]). As a result, a formation of the 
chemical spots is unlikely if there are hydrodynamic motions even 
with a weak shear. 

The current-driven diffusion in combination with other diffusion
mechanisms can be important for the surface chemistry of various 
types of stars. The mechanism considered can operate in various 
astrophysical bodies where the electric currents are non-vanishing. 
As it was noted, the current-driven diffusion leads to a formation 
of chemical spots only if the star has quiescent surface layers. 
That is the case, for instance, for white dwarfs and neutron stars. 
Observations detect strong magnetic fields in many neutron stars 
and, likely, topology of these fields should be rather complex with 
spot-like structures at the surface. As it was shown in our study, 
such magnetic structures can be responsible for the formation of 
element spots at the surface. A spot-like distribution of chemical 
elements can be important for the emission spectra, diffusive 
nuclear burning (see, e.\:g., [22, 23]), etc. Evolution of neutron 
stars is very complicated, particularly, in binary systems (see, 
e.\:g., [24]) and, as a result, a surface chemistry can be complicated
as well. Diffusion processes play an important role in this chemistry 
(see, e.\:g., [25, 26]) and can be the reason of chemical spots on 
the surface of these stars.   

Certainly, this type of diffusion may play an important role in the
surface chemistry of the so called Ap/Bp-stars. These stars have
a strong magnetic field [7] that magnetizes the atmospheric plasma
and produces a rapid Hall drift of electrons. Using Eq. (15), one can 
estimate the velocity of current-driven diffusion as
\begin{equation}
V_B \sim 1.1 \times 10^{-4} A^{-1/2}_i B_4^2 n_{15}^{-2} 
T_4^{3/2} \Lambda_{10} L_{B \; 10}^{-1} \;\;{\rm cm/s},
\end{equation}  
where $\Lambda_{10} = \Lambda/10$, $B_4 =B /10^4$ G, and $L_{B \; 10} 
= L_B /10^{10}$~cm. The velocity $V_B$ turns out to be sensitive to 
the field ($\propto B^2$) and, therefore, diffusion in a weak 
magnetic field requires a longer time to reach equilibrium. Since 
$B_4 \sim 1$, $T_4 \sim 1$, and $L_{B \;10} \sim 1$ are more or less 
typical values for Ap/Bp stars one can estimate that the timescale 
of spot formation in the atmosphere is shorter than the lifetime 
of such stars. Therefore, the current-driven diffusion can contribute 
to the generation of chemical structures in these stars. The 
conditions in Ap/Bp stars are also suitable for the propagation of 
compositional waves and, likely, such waves can be the reason of 
variations in atmospheric abundances of these stars. 

The considered mechanism can operate in laboratory plasma as well.
For instance, plasma adiabatically compressed and heated in 
experiments with explosives can reach very high values of the 
temperatures $T \sim 10^7$--$10^8$~K, number density $n \sim 10^{20}$--$10^{21}$~cm$^{-3}$, and magnetic field $B\sim 10^6$~G. In  multiple 
mirror experiments with the improved confinement (see, e.\:g., [27]), 
the number density is typically lower ($n \sim 10^{18}$~cm$^{-3}$) 
or even $\sim 10^{16}$~cm$^{-3}$ if $\text{CO}_{2}$ laser is used for 
heating. In such conditions, even impurities are usually strongly 
magnetized and $q \geq 1$. Nevertheless, the current-driven diffusion 
is still rather efficient and the diffusion velocity $V_B$ (Eq.~(16)) 
reaches the values $\sim 10^4$--$10^5$~cm/s. Correspondingly, the chemical 
structures can be generated in such plasma on a timescale of 
milliseconds. The considered diffusion mechanism can also operate 
in plasma of $\theta$-pinch. Such configurations are very suitable 
to study diffusion processes because of their long lifetime. Typical
number density and temperature are $\sim 10^{18}$~cm$^{-3}$ and
$10^7$--$10^8$~K, respectively. Plasma is essentially magnetized in
$\theta$-pinch since $q \sim 10^3$--$10^4$ (see, e.\:g., [1]) but,
nevertheless, there is enough time for generation of chemical 
spots because of sufficiently long lifetime. 

Our study reveals that a particular type of waves may exist in
multicomponent plasma in the presence of electric currents. These 
waves are slowly decaying and characterized by oscillations of the 
impurity number density alone. They exist only if the magnetic 
field is 
so strong that the magnetic pressure is greater than 
the gas pressure. Generally, the frequency of such waves turns 
out to be different for different impurities. This frequency is 
rather low and is determined mainly by a diffusion timescale. If 
$M=0$, it can be estimated as $\omega_I \sim k D_B /L \sim c_A^2 
\tau_i / A_i L \lambda$ where $\lambda = 2 \pi/k$ is the wavelength 
of waves. In astrophysical conditions, such waves can manifest 
themselves in the atmospheres of magnetic stars where the magnetic 
field is of the order of $10^4$ G and the number density and 
temperature are $10^{14}$ cm$^{-3}$ and $10^4$ K, respectively. If 
the lengthscale, $L$, and the wavelength, $\lambda$, are of the 
same order of magnitude (for instance, $\sim 10^{11}$ cm), then 
the period of compositional waves is $\sim 3 \times 10^3$ yrs. 
This is much shorter than the stellar lifetime and generation of 
such waves in the atmosperes should lead to spectral variability 
with the corresponding timescale. 

Compositional waves can occur in laboratory plasmas as well but 
their frequency is essentially higher. If $B \sim 10^5$ G, $n 
\sim 10^{15}$ cm$^{-3}$, $T \sim 10^6$~K, and $L \sim \lambda \sim 
10^2$~cm, than the period of compositional waves is $\sim 10^{-8}$ s. 
Note that this is only the order of magnitude estimate but 
frequencies of various impurities can differ essentially since the 
period of compositional waves depends on the sort of heavy ions.
In terrestrial conditions, the compositional waves also can manifest 
themselfes by oscillations in spectra. Note that these 
waves exist only if the magnetic pressure is greater than 
the gas pressure. The current-driven
diffusion can be important not only in plasma but in some conductive 
fluids if the magnetic field is sufficiently strong there. 

The author thanks the Russian
Academy of Sciences for financial support under the 
program OFN-15.

\newpage
\section*{REFERENCES}

\noindent
1. G. Vekshtein, {\it Reviews of Plasma Physics}, \textbf{15}, 1 (1987).\\
2. Y. Ren, M. Yamada, H. Ji, S. Gerhardt, and R. Kulsrud,
Phys. Rev. Lett., \textbf{101}, 5003 (2008).\\
3. G. Kagan and X.Z. Tang, Phys. Plasma, \textbf{107}, 50030 (2012). \\
4. T. Losseva, S. Popel, M.Y. Yu, and  J.X. Ma, Phys. Rev. E, \textbf{75}, 
6403 (2007). \\
5. T. Ott, and M. Bonitz, Phys. Rev. Lett., \textbf{107}, 50030 (2011). \\
6. K. Molvig, E. Vold, E. Dodd, and S. Wilks, Phys. Rev. Lett., \textbf{113},  9904 (2014). \\
7. V. Khokhlova, \textit{Sov. Sci. Rev}. (\textit{Sec. E: 
Astrophysics and Space Phys. Reviews}), \textbf{4}, 99 (1985). \\
8. G. Mathys and S. Hubrig, Astron. Astrophys., \textbf{293}, 810 (1995).\\
9. O. Kochukhov, IAU Symposium, \textbf{224}, 433 (2004). \\
10. L. Spitzer, {\it Physical Processes in the Interstellar
Medium} (New York: Wiley. 1978) \\
11. G. Vekshtein, D. Riutov, and P. Chebotaev, Soviet J. Plasma Phys., \textbf{1}, 220 (1975).\\
12. R.J. Tayler, Monthly Notices Royal Astron. Soc., \textbf{161}, 365 (1973). \\
13. A. Bonanno and V. Urpin, Astron. Astrophys., \textbf{477}, 35 (2008). \\
14. A. Bonanno and V. Urpin, Astron. Astrophys., \textbf{488}, 1 (2008). \\
15. V. Urpin and K. van Riper, Astrophys. J., \textbf{411}, L87 (1993). \\ 
16. S. Braginskii, {\it Reviews of Plasma Physics}. 
(Consultants Bureau, New York, 1965) \textbf{1}, 205. \\
17. V. Urpin, Astrophys. Space Sci., \textbf{79}, 11 (1981). \\
18. V. Urpin, Astron. Nachr., \textbf{336}, 266 (2015). \\
19. A.B. Mikhailovskii, J. Lominadze, A. Churikov, and V. Pustovitov,
Plasma Phys. Rep., \textbf{35}, 273 (2009). \\
20. C. Bejerano, D. Gomez, and A. Brandenburg, Astrophys. J., \textbf{737}, 62 (2011). \\
21. V. Urpin and G. R\"udiger, Astron. Astrophys., \textbf{437}, 23 (2005). \\
22. E. Brown, L. Bildsten, and P. Chang, Astrophys. J., \textbf{574}, 920 (2002). \\
23. P. Chang and L. Bildsten, Astrophys. J., \textbf{605}, 830 (2004). \\
24. V. Urpin, D. Konenkov, and U. Geppert, Monthly Notices Royal Astron. Soc., \textbf{299}, 73 (1998). \\
25. E. Brown, L. Bildsten, and P. Chang, Astrophys. J., \textbf{574}, 920 (2002). \\
26. Z. Medin, and A. Cumming, Astrophys. J., \textbf{783}, 3 (2014). \\
27. A. Burdakov, A. Ivanov, and E. Kruglyakov, Plasma Phys. Controlled Fusion, \textbf{521} 4026 (2010). \\

\end{document}